\documentclass[twocolumn,aps,prb,showpacs,amsfonts,amssymb,amsmath]{revtex4}

\usepackage{amsfonts}
\usepackage{amsmath}
\usepackage{amssymb}
\usepackage{graphicx}
\begin{document}

\title{Measuring the phonon-assisted spectral function by using a non-equilibrium three-terminal
single-molecular device}

\author{Juntao Song$^{1}$, Qing-feng Sun$^{1,\ast}$, Jinhua Gao$^{1,2}$,
and X.C. Xie$^{1,2}$} \affiliation{$^1$Beijing National Lab for
Condensed Matter Physics and Institute of Physics, Chinese Academy
of Sciences, Beijing
100080, China \\
$^2$Department of Physics, Oklahoma State University, Stillwater
Oklahoma, 74078 USA}

\date{\today}

\begin{abstract}
The electron transport through a three-terminal single-molecular
transistor (SMT) is theoretically studied. We find that the
differential conductance of the third and weakly coupled terminal
versus its voltage matches well with the spectral function versus
the energy when certain conditions are met. Particularly, this
excellent matching is maintained even for complicated structure of
the phonon-assisted side peaks. Thus, this device offers an
experimental approach to explore the shape of the phonon-assisted
spectral function in detail. In addition we discuss the conditions
of a perfect matching. The results show that at low temperatures the
matching survives regardless of the bias and the energy levels of
the SMT. However, at high temperatures, the matching is destroyed.
\end{abstract}

\pacs{73.63.Kv, 71.38.-k, 85.65.+h}

\maketitle

\section{Introduction}

In the past decade, transport properties of single-molecule
transistors (SMT) have attracted great attention due to the
potential application for the new generation of electron devices.
Because of the intrinsic vibrational freedom in molecules, the
molecular electronic transistor also provides a new opportunity for
exploring the vibration-electronic (i.e. electron-phonon)
interactions at single molecule level. The electron-phonon
interaction in a SMT leads to some interesting effects, such as
phonon-assisted tunneling, the red shift of SMT energy levels, and
generation of the thermal energy. Such features are interesting and
have been extensively investigated both experimentally and
theoretically in recent years. Park {\sl et al.}~\cite{H. Park}
experimentally studied the current-bias (I-V) characteristic of an
individual C$_{60}$ molecule connected to gold electrodes and have
observed the obvious phonon-assisted tunneling sub-steps in the I-V
curves. Later the current of a suspended individual single-wall
nanotube device is measured and two phonon-assisted sub-peaks on the
two sides of the main resonant peak are clearly visible in the
differential conductance versus the gate voltage, which is due to
the radial breathing phonon mode.~\cite{B. J. Leroy1,B. J. Leroy2}
Very recently, also in the device of a suspended single-wall
nanotube but with much lower temperatures, the higher order
phonon-assisted sub-steps on the I-V curves have been experimentally
demonstrated by Sapmaz {\sl et al.}~\cite{S. Sapmaz}

On the theoretical side, there have also been a large amount of
studies on the quantum transport behavior of a SMT or a quantum dot
(QD) coupled to local phonon modes.~\cite{N. S. Wingreen,M. Keil,D.
M.-T. Kuo,J. X. Zhu,K. Flensberg,Z. Z. Chen,A. Mitra} About 15 years
ago, Wingreen {\sl et al.} studied the electron transport through a
QD coupled to the phonon modes by combining the scattering theory
and the Green's function method, and the phonon-induce transmission
sidebands were found.~\cite{N. S. Wingreen} Using the real-time
renormalization-group method, Keil {\sl et al.}~\cite{M. Keil} have
investigated the quantum transport phenomena through coupled QDs
with a phonon bath, and a solution for stationary current is
obtained. In addition, the shot-noise spectroscopy of the current of
a SMT having a local phonon mode is reported by Zhu and
Balatsky.\cite{J. X. Zhu} Since the current, the conductance, the
shot-noise, {\sl etc.}, are all closely related to the local
electronic spectral functions $A(\omega)$, the spectral function has
also been extensively studied. The spectral function is found to be
strongly dependent on the positions of the SMT (or QD) energy levels
and the tunneling strengths $\Gamma$ between the leads and the SMT
(or QD).\cite{K. Flensberg} On the other hand, by using a different
approximation, Chen {\sl et al.} also investigated the spectral
function and the current through a SMT,\cite{Z. Z. Chen} and the
spectral function in their results exhibits intriguing features. At
low temperatures and under the condition that the energy level is
near the Fermi surface, the side peaks in the spectral function are
clearly non-Lorenzian in shape. The side peaks on one side changes
gradually while the other side change abruptly with changing of
energy. These are quite different comparing with the previous
results. Due to the importance of the spectral function $A(\omega)$
as well as that the spectral function of a SMT coupled to a phonon
mode is not well understood (i.e. the results of $A(\omega)$ are
qualitatively different for using different approximations), thus it
is quite beneficial to design an experimental set-up to directly
measure the spectral function of a SMT.

Recently, some studies have investigated the Kondo effect in QDs
coupled to the three terminals, in which the third terminal acts as
an exploring tip to measure the Kondo peaks in the spectral
function.\cite{Kondo1,Kondo2} Can the intriguing characteristics of
the phonon-assisted side peaks in the spectral function also be
explored in a three-terminal set-up? It is the purpose of this work
to theoretically analyze the feasibility of this scheme. We consider
the system of either SMT or QD coupled to three leads and a local
phonon mode. Here the third lead is introduced as an exploring tip.
We find that the spectral function quite often matches the
differential conductance of the third lead. If the third lead is
weakly coupled and the temperature $T$ is low ($k_B T \ll
\hbar\omega_0$ with $\omega_0$ being the phonon frequency), this
matching is almost perfect (including the abrupt changes associated
with the phonon-assisted side peaks), so the spectral function, in
particularly the phonon-assisted side peaks, can be directly
detected using the differential conductance. On the other hand, if
the coupling of the third lead is large but still at low
temperature, the matching is in qualitative agreement. However, at
high temperature, the matching is destroyed even for weakly-coupled
third lead.

The rest of this paper is organized as follows. We introduce the
model in Sec. II and derive formulas of the spectral function and
the differential conductance in Sec. III. In Sec. IV, we show the
numerical results and present discussions of those results.
Finally, a brief summary is given in Sec. V.

\section{MODEL HAMILTONIAN}
The device under our consideration is illustrated in Fig.1. It
consists of a SMT or QD connected to three metallic leads. An
electron in the SMT is also coupled to a single phonon mode. Due
to the big energy gap between two quantum levels in the SMT, only
one relevant quantum level is considered. The Hamiltonian of the
device is written as:
\begin{equation}
H=H_{Leads}+H_{ph}+H_{D}+H_{T}.
\end{equation}
The first two terms are, respectively, the Hamiltonian for
electrons in the three leads and the Hamiltonian for the phonon
part,
\begin{eqnarray}
&&H_{leads}=\sum_{\alpha,k}{\varepsilon_{\alpha k}c_{\alpha
k}^\dagger c_{\alpha k}},
\\
&&H_{ph}=\omega_0{b^\dagger b}.
\end{eqnarray}
Here $\alpha=L$, $R$, and $3$ respectively represent the left, the
right, and the third leads, and $c_{\alpha k}^\dagger (c_{\alpha
k})$ creates (annihilates) an electron with the energy
$\varepsilon_{\alpha k}$ in the lead $\alpha$. Analogously,
$b^\dagger (b)$ is the phonon creation (annihilation) operator and
$\omega_0$ is the vibrational frequency of the phonon. The third
term in Eq.(1) is
\begin{eqnarray}
H_{D}=[\varepsilon_0+\lambda (b^\dagger+ b)]{d^\dagger d},
\end{eqnarray}
where $d^\dagger (d)$ is the electron creation (annihilation)
operator in the SMT with the energy level $\varepsilon_0$, and
$\lambda$ describes the coupling strength between the SMT and the
local phonon mode. The last term in Eq.(1) descries the tunnelling
coupling between the SMT and the three leads,
\begin{eqnarray}
H_{T}=\sum_{\alpha,k}{[V_{\alpha k} c_{\alpha k}^\dagger d+H.c]}.
\end{eqnarray}
It is often useful to take a canonical transformation with:\cite{Z.
Z. Chen,G. D. Mahan} $ \overline{H}=e^{s}H e^{-s}$ and $s=(\lambda
/\omega_0) (b^\dagger- b){d^\dagger d}£¬ $. Under this canonical
transformation the Hamiltonian (1) varies into
\begin{eqnarray}
\overline{H}=\overline{H}_{el}+\overline{H}_{ph},
\end{eqnarray}
where
\begin{eqnarray}
\overline{H}_{el}=\sum_{\alpha,k}{\varepsilon_{\alpha k}c_{\alpha
k}^\dagger c_{\alpha k}}+\overline{\varepsilon}_0d^\dagger
d+\sum_{\alpha,k}{[\overline{V}_{\alpha k} c_{\alpha k}^\dagger
d+H.c]},
\end{eqnarray}
and
\begin{eqnarray}
\overline{H}_{ph}=\omega_0b^\dagger b.
\end{eqnarray}
Obviously, due to the electron-phonon interaction, the energy
level ${\varepsilon}_0$ of the SMT is renormalized to
$\overline{\varepsilon}_0=\varepsilon_0-g\omega_0$ with
$g=(\lambda/\omega_0)^2$, and the tunnelling matrix element $V_k$
is varied into $\overline{V}_k=V_kX$ where
$X=\exp{[-(\lambda/\omega_0)(b^\dagger+b)]}$. Up till now, no
approximation has been made, the Hamiltonian (6) is completely
equivalent to the Hamiltonian (1).\cite{G. D. Mahan}

\section{THE SPECTRAL FUNCTION AND THE DIFFERENTIAL CONDUCTANCE}

In this section, we calculate the spectral function $A(\omega)$ of
the SMT and the differential conductance of the third lead. From
the results by Meir and Wingreen, the spectral function
$A(\omega)$ and the current can be represented by the Green
functions of the SMT as:\cite{Meir,A. P. Jauho}
\begin{eqnarray}
A(\omega)=i[G^>(\omega)-G^<(\omega)]=i[G^r(\omega)-G^a(\omega)].
\end{eqnarray}
\begin{eqnarray}
 J_{\alpha}
 = \frac{e}{\hbar}\int\frac{d\omega}{2\pi}Tr\{\Gamma_{\alpha}
 [iG^<(\omega)+A(\omega)f_{\alpha}(\omega)]\}.
\end{eqnarray}
Here $f_{\alpha}(\omega)= 1/\{exp[(\omega-\mu_{\alpha})/k_BT]+1\}$
is the Fermi distribution function with the chemical potential
$\mu_{\alpha}$, $\Gamma_{\alpha}(\omega)=2\pi \sum_k |V_{\alpha
k}|^2\delta(\varepsilon_{\alpha k}-\omega)$ describes the coupling
strength between the lead $\alpha$ and the SMT, and $G^{<,>,r,a}$
are the standard lesser, greater, retarded, advanced Green
functions.\cite{A. P. Jauho,H. Huag} Because of the existence of the
electron-phonon interaction, it is difficult to directly solve these
Green functions from the equation of motion technique or the Dyson
equations. As it is done in some previous papers,~\cite{Z. Z. Chen}
here we take the same approximation to replace the operator $X$ and
$X^\dagger$ in Hamiltonian (6) with their expectation value $\langle
X \rangle=\langle X^\dagger\rangle=\exp{[-g(N_{ph}+1/2)]}$, where
$N_{ph}=1/[\exp(\beta\omega_0)-1]$ is the phonon
population.\cite{addnote} This approximation is valid when the
tunneling strengths $\Gamma_{\alpha }$ are smaller than the
electron-phonon interaction, i.e, $\Gamma_{\alpha }\ll \lambda$.
After this approximation, the Hamiltonian (6) is decoupled into two
independent parts, electronic part and phonon part. Next, we also
need to decouple the Green functions. In many previous
papers,~\cite{J. X. Zhu} they decouple the retarded (advanced) Green
functions $G^{r,a}$ directly. However, such decoupling has some
defects as pointed out in a recent work by Chen {\sl et
al}..~\cite{Z. Z. Chen} Here we employ the decoupling method in the
Ref.(\cite{Z. Z. Chen}), to directly decouple the lesser and greater
Green functions $G^{<,>}$ instead of the retarded and advanced Green
functions $G^{r,a}$. After the decoupling, the lesser and greater
Green functions $G^{<,>}$ are:
\begin{eqnarray}
G^{<}(t,t') &= & \overline{G}^{<}(t,t')\ e^{-\Phi(t'-t)},\\
G^{>}(t,t') &= & \overline{G}^{>}(t,t')\ e^{-\Phi(t-t')},
\end{eqnarray}
where $\overline{G}^{<,>}(t,t')$ are the Green functions of the
Hamiltonian $\overline{H}_{el}$, and $\Phi(t)$ is
\begin{eqnarray}
 \Phi(t) = g\left[ N_{ph} (1-e^{i\omega_0 t})
  +(N_{ph}+1)(1-e^{-i\omega_0 t}) \right].
\end{eqnarray}
Using the identity~\cite{G. D. Mahan}
$e^{z\cos\theta}=\sum^{n=+\infty}
_{n=-\infty}{I_n(z)}e^{in\theta}$, the greater and lesser Green
functions can be expanded as
\begin{eqnarray}
G^{<}(\omega) =\sum^{+\infty}
_{n=-\infty}{B_n}\overline{G}^{<}(\omega+n\omega_0),\\
G^{>}(\omega) =\sum^{+\infty}
_{n=-\infty}{B_n}\overline{G}^{>}(\omega-n\omega_0),
\end{eqnarray}
where the coefficients $B_n=e^{-g(2N_{ph}+1)}e^{n\omega_0/2k_B
T}I_n\big(2g\sqrt{N_{ph}(N_{ph}+1)}\ \big)$ and $I_n(z)$ is the
nth Bessel function of complex argument. Thus we can rewrite the
spectral function $A(\omega)$ as
\begin{eqnarray}
 A(\omega)&= &i[G^>(\omega)-G^<(\omega)]\nonumber\\
&= &\sum^{+\infty}
_{n=-\infty}i{B_n}[\overline{G}^{>}(\omega-n\omega_0)-\overline{G}^{<}(\omega+n\omega_0)].
\end{eqnarray}
Following the standard derivation,~\cite{G. D. Mahan,A. P.
Jauho,H. Huag} the self-energies $\overline{\Sigma}$ of the
coupling to the leads for the Hamiltonian $\overline{H}_{el}$ can
be easily obtained as:
\begin{eqnarray}
\overline{\Sigma}^{r(a)}(\omega) &
 = &\sum_{\alpha,k}{|\overline{V}_{\alpha k}|^2g^{r(a)}_{\alpha
k}}(\omega)=\sum_\alpha[\overline{\Lambda}_\alpha(\omega) \mp
\frac{i}{2}\overline{\Gamma}_\alpha(\omega)], \\
 \overline{\Sigma}^{<}(\omega)
 &= &\sum_{\alpha,k}{|\overline{V}_{\alpha k}|^2 g^{<}_{\alpha k}}(\omega)
=i\sum_\alpha\overline{\Gamma}_\alpha(\omega)f_\alpha(\omega),\\
\overline{\Sigma}^{>}(\omega)
 &= &\sum_{\alpha,k}{|\overline{V}_{\alpha k}|^2 g^{>}_{\alpha k}}(\omega)
=-i\sum_\alpha\overline{\Gamma}_\alpha(\omega)[1-f_\alpha(\omega)],
\end{eqnarray}
where $\overline{\Gamma}_\alpha=\Gamma_\alpha \exp{[-g(2N_{ph}+1)]}$
since the tunneling elements $V_{\alpha k}$ have been amended by
electron-phonon interaction. To take the wideband limit,~\cite{A. P.
Jauho,H. Huag} i.e. to assume that $\Gamma_{\alpha}$ then
$\overline{\Gamma}_{\alpha}$ are independent with the energy
$\omega$, the above self-energies reduce into:
\begin{eqnarray}
\overline{\Sigma}^{r(a)}
 &=
 &\mp\frac{i}{2}(\overline{\Gamma}_L+\overline{\Gamma}_R+\overline{\Gamma}_3),\\
 \overline{\Sigma}^{<}(\omega)
 &= &i\sum_\alpha{\overline{\Gamma}_\alpha f_\alpha(\omega)},\\
 \overline{\Sigma}^{>}(\omega)
 &= &-i\sum_\alpha{\overline{\Gamma}_\alpha [1-f_\alpha(\omega)]}.
\end{eqnarray}
By using these self-energies, the dressed retarded (advanced)
Green function $\overline{G}^{r(a)}$, then the dressed lesser and
greater Green functions $\overline{G}^{<,>}$ can be readily
obtained from Dyson equations and Keldysh equations:
\begin{eqnarray}
\overline{G}^{r(a)}(\omega)
 &=
 &[{\overline{g}^{r(a)}(\omega)-\overline{\Sigma}^{r(a)}(\omega)}]^{-1},\\
 \overline{G}^{<}(\omega)
 &=
 &\overline{G}^{r}(\omega)\overline{\Sigma}^{<}(\omega)\overline{G}^{a}(\omega)
=i\overline{f}(\omega)\overline{A}(\omega),\\
\overline{G}^{>}(\omega)
 &= &\overline{G}^{r}(\omega)\overline{\Sigma}^{>}(\omega)\overline{G}^{a}(\omega)
=-i[1-\overline{f}(\omega)]\overline{A}(\omega),
\end{eqnarray}
where
\begin{eqnarray}
\overline{A}(\omega)=\frac{\overline{\Gamma}_L+\overline{\Gamma}_R+\overline{\Gamma}_3}
{(\omega-\overline{\varepsilon}_0)^2+(\overline{\Gamma}_L+\overline{\Gamma}_R+\overline{\Gamma}_3)^2/4},
\end{eqnarray}
and
\begin{eqnarray}
\overline{f}(\omega)=\frac{\overline{\Gamma}_L
f_L(\omega)+\overline{\Gamma}_Rf_R(\omega)+\overline{\Gamma}_3f_3(\omega)}
{\overline{\Gamma}_L+\overline{\Gamma}_R+\overline{\Gamma}_3}.
\end{eqnarray}
After solving $\overline{G}^{<,>}$, the lesser and greater Green
functions $G^{<,>}(\omega)$, the electronic spectral function
$A(\omega)$ of the SMT, and then the current can be calculated
from the above Eqs. (9), (10), (14), and (15), straightforwardly.

At last, the differential conductance $G_3$ of the third terminal
can be acquired by performing $G_3=\partial J_3/\partial V_3$
\begin{eqnarray}
G_3&=&\sum^{+\infty}_{n=-\infty}\frac{e^2\Gamma_3}{\hbar
kT} B_n\int\frac{d\omega}{2\pi}\big\{f_3(\omega)[1-f_3(\omega)]\nonumber\\
&&\times\big[\overline{A}(\omega_2)\overline{f}(\omega_2)
+\overline{A}(\omega_1)[1-\overline{f}(\omega_1)]\big] \nonumber\\
&&-c [1-f_3(\omega)]f_3(\omega_2)[1-f_3(\omega_2)]\overline{A}(\omega_2)\nonumber\\
&&-c
f_3(\omega)f_3(\omega_1)[1-f_3(\omega_1)]\overline{A}(\omega_1)\big\},
\end{eqnarray}
where $\omega_1=\omega-n\omega_0$, $\omega_2=\omega+n\omega_0$,
and $c =\overline{\Gamma}_3/
(\overline{\Gamma}_L+\overline{\Gamma}_R+\overline{\Gamma}_3)$.

\section{Numerical Results and Discussions}

In this section, we study numerically the spectral function
$A(\omega)$ and the differential conductance $G_3$. In the numerical
calculation, the coupling strengths $\Gamma_{L/R}$ between the
left/right lead and the SMT is set to be unity
($\Gamma_L=\Gamma_R\equiv\Gamma =1$), as an energy unit. The main
purpose in the present work is to study whether the curve of the
spectral function $A(\omega)$ versus the energy $\omega$ can map
into the curve $G_3$-$V_3$, i.e. whether the intriguing
phonon-assisted side peaks in the spectral function $A(\omega)$ can
be observed by measuring the conductance $G_3$ of the third lead.
First, let us show the spectral function of the two-terminal SMT
with $\Gamma_3=0$. Note that this spectral function $A(\omega)$ is
the object of our study. When the SMT is coupled to the phonon mode,
one main characteristic of $A(\omega)$ is the appearance of the
phonon-assisted side peaks. At the zero bias case ($\mu_L=\mu_R=0$)
and the renormalized level $\overline{\varepsilon}_0 =0$, the side
peaks are non-Lorenzian in shape, in which one side of the side
peaks still looks like the Lorenzian form but the other side drops
abruptly (see Fig.2). When $\overline{\varepsilon}_0$ (i.e
$|\overline{\varepsilon}_0|/\Gamma \gg 0$) is far away from the
chemical potentials $\mu_L,\mu_R$, the side peaks are asymmetry on
the two sides of the main peak, and the side peaks disappear on one
side. Furthermore, with a non-zero bias $V$ ($V=\mu_L-\mu_R$) or
raising temperature $T$, these phonon-assisted side peaks exhibit
more complex profiles. These characteristics of the spectral
function $A(\omega)$ have been found in a previous study.\cite{Z. Z.
Chen} Our goal here is to propose a scheme to measure the spectral
function $A(\omega)$ by using an extra third lead.

We first study the zero bias case ($\mu_L=\mu_R=0$) with a
weakly-coupled third lead ($\Gamma_3=0.01$). Fig.2 shows the
differential conductance $G_3$ versus the voltage $V_3$ of the third
lead for different renormalized level $\overline{\varepsilon}_0$.
For comparison, the spectral function $A(\omega)$ versus the energy
$\omega$ for the two-terminal SMT device with $\Gamma_3=0$ is also
shown in Fig.2. When $\overline{\varepsilon}_0=0$. Besides the main
resonant tunneling peak, some extra phonon-assisted side peaks
emerge in the curve $G_3$-$V_3$. The main peak is Lorenzian, but the
side peaks exhibit non-Lorenzian characteristic. On the one side of
the side peaks, the conductance $G_3$ falls abruptly from top to
valley. In particular, we find that the curve of the conductance
$G_3$ versus $V_3$ is in an excellent agreement with the curve of
the spectral function $A(\omega)$ versus the energy $\omega$. Not
only are their side peaks located at the same positions, but also
they have the same profiles. Even the abrupt drops overlap
perfectly. Thus, in this case by measuring the differential
conductance $G_3$, one obtains all information on the spectral
function $A(\omega)$. Increasing $\overline{\varepsilon}_0$ to $1$,
the side peaks in the curve of $G_3$-$V_3$ are distributed
asymmetrically on two sides of the main resonant peak. The right
side peak is higher than the corresponding left side peak as shown
in Fig.2. However the curve of $G_3$-$V_3$ is still in an excellent
matching with the spectral functions $A(\omega)$ versus $\omega$,
even the complex structure of the first right side peak (about at
$V_3 =6$) matches quite well. When $\overline{\varepsilon}_0$ is
furthermore enhanced from $1$ to $3$, all side peaks are on the
right of the main peak and all peaks are Lorenzian. Similarly the
excellent agreement between the differential conductance and the
spectral function are still maintained. Combining above results we
find that at the zero bias case, the curve $G_3$-$V_3$ is in
excellent agreement with the curve $A(\omega)$-$\omega$ regardless
of the value of the level $\overline{\varepsilon}_0$ (i.e.
$\varepsilon_0$). The reason lies in the fact that for the
weakly-coupled lead, the transmission probability of the incoming
electron with energy $\omega$ is mainly determined by the local
spectral function $A(\omega)$.

Next, we study the case with a finite bias $V$ ($V=\mu_L-\mu_R$).
Fig.3 shows the differential conductance $G_3$ and the spectral
function $A(\omega)$ for $\mu_L = -\mu_R =3$. In the finite bias
$V=6$ and the renormalized level $\overline{\varepsilon}_0=0$, the
phonon-assisted side peaks in the conductance $G_3$ are
symmetrically distributed on two sides of the main resonant peak,
and the form of the side peaks are Lorenzian which is in contrast to
zero bias case with non-Lorenzian side peaks (see Fig.2). However,
the characteristics of the spectral function are still reflected
perfectly by the differential conductance, although the spectral
functions for the zero bias and the non-zero bias have a large
difference. When $\overline{\varepsilon}_0$ is reduced to $-3$, the
left phonon-assisted side peak is obviously higher than the
corresponding right side peak, the first left side peak becomes
sharp, and the non-Lorenzian characteristic emerges again in the
first right side peak. Although the spectral function $A(\omega)$ is
so complex now, the curve of $G_3$-$V_3$ closely follows that of the
spectral function, including the detail structure. With decreasing
$\overline{\varepsilon}_0$ further, e.g. $\overline{\varepsilon}_0
=-6$, the conductance $G_3$ has a large change. In the present case,
the side peaks of the right hand disappear completely and the side
peaks only exit on the left hand of the main peak. We find that the
curve $G_3$-$V_3$ completely matches the curve $A(\omega)$-$\omega$.
In fact, for any bias $V$ and any level $\overline{\varepsilon}_0$
(i.e. $\varepsilon_0$), the curves $G_3$-$V_3$ and
$A(\omega)$-$\omega$ overlap perfectly regardless of complexity of
the curve $A(\omega)$-$\omega$. This means that by measuring the
differential conductance $G_3$ of the third lead, the spectral
function $A(\omega)$, including the intriguing characteristics due
to coupled to the phonon mode, can be directly observed.

In the above numerical investigation, the coupling $\Gamma_3$
between the SMT and the third lead is set to be rather weak
($\Gamma_3=0.01$), and the temperature $T$ is kept at very low
($k_BT=0.02$). For a strongly-coupled third lead and at high
temperature, will the excellent agreement between the differential
conductance $G_3$ and the spectral function $A(\omega)$ still
survive? In this paragraph, we study the effect of the coupling
strength $\Gamma_3$. The temperature effect is addressed in the next
paragraph. Fig.4(a) and (b) show the conductance $G_3$ and the
spectral function $A(\omega)$ for different coupling strength
$\Gamma_3$. With increasing $\Gamma_3$, the level in the SMT is
widen, and the conductance $G_3/G_0$ [$G_0=(2e^2/\hbar)
(\Gamma_3/\Gamma)$] is overall reduced. To make a better comparison
of the two curves $G_3$-$V_3$ and $A(\omega)$-$\omega$, we replace
the conductance unit $G_0$ by an integral weighting factor $G'_0$
and Fig.4 (c) and (d) show $G_3/G'_0$ versus $V_3$, here $G'_0$ is
determined by the equation $\int\frac{dV_3}{2\pi}\frac{G_3}{G'_0} =
\int\frac{d\omega}{2\pi}\frac{A(\omega)}{A_0}$. The results are as
following: (i) When the third lead is weakly-coupled with a small
$\Gamma_3$ ($\Gamma_3=0.01$), the differential conductance $G_3$ and
the spectral function $A(\omega)$ map into each other perfectly, as
discussed before. (ii) With increasing coupling strength $\Gamma_3$,
the differential conductance deviates from the spectral function
gradually. When $\Gamma_3$ is in the same order of $\Gamma$ (e.g
$\Gamma_3 =0.4$ or $1$), the peak of the conductance $G_3$ becomes
lower and wider than that of the spectral function (see Fig.4). But
overall the conductance $G_3$ still shows the similar profile of the
spectral function, including the abrupt drop on the phonon-assisted
side peak. In other words, the curve $G_3$-$V_3$ is still
qualitatively the same as the curve $A(\omega)$-$\omega$ when
$\Gamma_3 \sim \Gamma$. Therefore, the probing lead is not
necessarily need to be weakly coupled to the SMT, and the device can
still work at $\Gamma_3 \sim \Gamma$. (iii) When $\Gamma_3$ is much
larger than $\Gamma$ (e.g. $\Gamma_3=4\Gamma$ or more), the side
peaks in the conductance $G_3$ fade away, and the curves of
$G_3$-$V_3$ and $A(\omega)$-$\omega$ have a large discrepancy. Since
the properties [including the spectral function $A(\omega)$] of SMT
are remarkably influenced when the coupling between the third lead
(i.e. exploring terminal) and the SMT is very strong, the
differential conductance can no longer exhibit the characteristics
of the real spectral function. From Fig.4, it can be concluded that
the qualitative agreement between the curves of $G_3$-$V_3$ and
$A(\omega)$-$\omega$ is destroyed when $\Gamma_3
>2(\Gamma_L+\Gamma_R)$.

Now let us consider how temperature influence the relationship
between the spectral function and the differential conductance. As
displayed in Fig.5a, at low temperature ($T=0.05\Gamma$) the curve
of the spectral function overlaps perfectly with that of the
differential conductance, as being discussed before. When
temperature is raised to $0.1\Gamma$, although the peaks of the
differential conductance are slightly lower (about a few percent)
and wider than those of the spectral function, they still agree not
only in their positions, but also in the shape of the peaks and the
detail structure of the phonon-assisted side peaks (see Fig.5b).
With further increasing temperature $T$, the deviation between the
conductance $G_3$ and the spectral function $A(\omega)$ is more
noticeable. When $T=0.5\Gamma$, the peak heights of the conductance
decrease to half of those of the spectral function. In particular,
the shapes of the side peaks in the conductance and in the spectral
function are clearly in disparity. The shapes of the side peaks of
the spectral function are still asymmetric and the change is quite
abrupt on one side of the side peak, but the shapes of the side
peaks of the differential conductance are Lorenzian and symmetric
(see Fig.5c). While the temperature is equal to or larger than
$\Gamma$, all side peaks of the differential conductance fade away,
and the differential conductance is no longer providing any
information on the spectral function (see Fig.5d). Therefore, it is
feasible only at the low temperature to observe the spectral
function by measuring the differential conductance of the third
terminal. In the Fig.5, we choose zero bias and the renormalized
level $\overline{\varepsilon}_0$ located at zero. In fact, all
conclusions remain for non-zero bias and any value of
$\overline{\varepsilon}_0$.

Unlike the coupling strength $\Gamma_3$, the temperature strongly
influences the comparability of the curves $G_3$-$V_3$ and
$A(\omega)$-$\omega$. When temperature is low enough ($k_BT \ll
\Gamma, \hbar\omega_0$), there is a well-defined boundary for the
occupied states and empty states in the exploring terminal. With a
change of the terminal voltage $V_3$, the change of incident
electrons concentrate at a very small energy region, thus the
differential conductance gives an excellent mapping of the spectral
function. On the other hand, this well-defined boundary for the
occupied and empty states is destroyed when the temperature $k_B T
\sim \Gamma$, so that the conductance $G_3$ and the spectral
function $A(\omega)$ have a large discrepancy. Naturally, if the
third lead can be individually fixed at low temperature, then the
spectral function can still be obtained from the differential
conductance of the third lead regardless of the temperature in other
parts of the system. In fact, the situation with low temperature for
the third lead is always comparable with the above discussed low
temperature regime. Let us discussed the realizability of the low
temperature condition $k_BT \ll \Gamma, \hbar\omega_0$ in the
present technology. In an experiment, the characteristic frequency
of phonon is about from $5meV$ to $50meV$,~\cite{B. J. Leroy1,B. J.
Leroy2} and the coupling strength $\Gamma$ is usually in the order
of $100\mu eV$. But the temperature can reach $50mK$ in the present
technology. So it should be achievable for the condition $k_BT \ll
\Gamma, \hbar\omega_0$. In addition, in experiments the third
weakly-coupled lead can be a probing STM tip, and then the coupling
strength can be easily controlled by adjusting the distance between
the STM tip and the SMT.

At end, we make one more comment. Since we have used the same
approximation as in Ref.(\cite{Z. Z. Chen}), the spectral function
$A(\omega)$ is completely same as in their work. If to take a
different approximation [e.g. as in Ref.(\cite{K. Flensberg})],
the spectral function $A(\omega)$, in particular the shape of the
phonon-assisted side peaks, perhaps may vary somewhat. However,
the perfect matching for the curves $G_3$-$V_3$ and
$A(\omega)$-$\omega$ still maintain as long as at low temperature
and weak coupling conditions are met.

\section{CONCLUSIONS}

In summary, we study the transport behaviors of the three terminal
SMT device coupled to a phonon mode. It is found that the
intriguing characteristic of the phonon-assisted side peaks in the
spectral function versus the energy can be directly observed from
the differential conductance versus the voltage of the third
weakly-coupled lead. In particular, not only the positions but
also the shapes and the detailed structure of the side peaks of
the spectral function can be perfectly mapped into the conductance
if certain experimental conditions are met. Moreover we determine
the conditions for this perfect mapping. The results exhibit that
this mapping is excellent at low temperature regardless of the
bias and the level of the SMT. The mapping is destroyed at high
temperature.

\section{ACKNOWLEDGMENTS}
We thank Jianing Zhuang and Haijun Zhang for their helpful
discussions. This work is supported by NSF-China under Grant Nos.
90303016, 10474125, and 10525418, US-DOE under Grant No.
DE-FG02-04ER46124, and NSF under CCF-052473.

\newpage

\begin{figure}
\includegraphics[bb=9mm 9mm 101mm 80mm, width=9cm,totalheight=6.5cm,clip=]{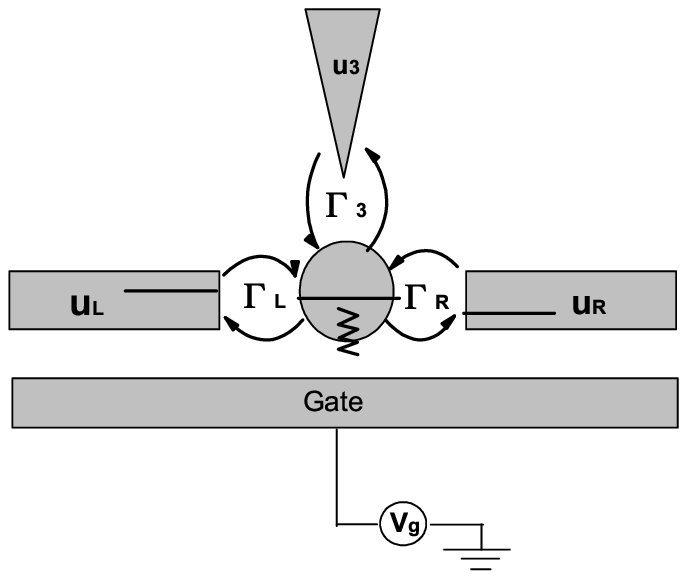}
\caption{ Schematic diagram of a SMT coupled to three leads with
tunnelling strength $\Gamma_{\alpha}$ and the electron in the SMT is
also coupled to a single-phonon mode. A gate electrode is
capacitively attached to the SMT to tune the energy level of SMT.}
\end{figure}

\begin{figure}
\includegraphics[bb=-1mm 9mm 121mm 85mm, width=8cm,totalheight=6.5cm,clip=]{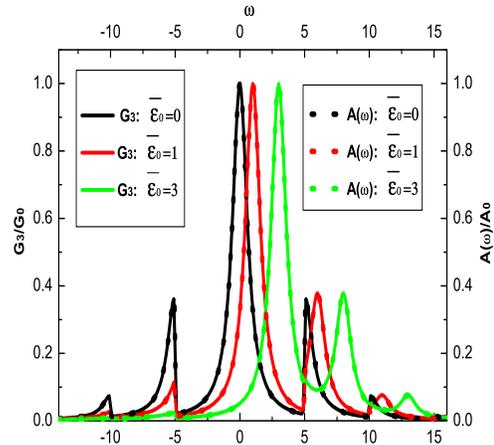}
\caption{(Color online) The dimensionless spectral function
$A(\omega)/A_0$ vs. the energy $\omega$ at $\Gamma_3=0$ and the
dimensionless differential conductance $G_3/G_0$ of the third lead
vs. the voltage $V_3$ at $\Gamma_3=0.01$ for different
$\overline{\varepsilon}_0$. Other parameters are taken as:
$u_L=u_R=0$, $\omega_0=5$, $T=0.02$, and $\lambda=3$. The unites
$A_0$ and $G_0$ are equal to $2/\Gamma$ and
$(2e^2/\hbar)(\Gamma_3/\Gamma)$ respectively. Notice that the three
dotted curves almost overlap perfectly with the three solid curves
so that they almost cannot be seen in the figure.}
\end{figure}

\begin{figure}
\includegraphics[bb=-1mm 9mm 121mm 108mm, width=8cm,totalheight=6.5cm,clip=]{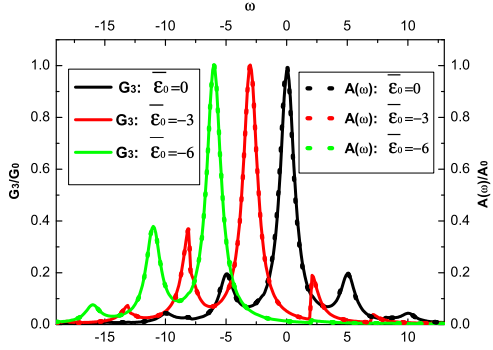}
\caption{(Color online) The dimensionless spectral function
$A(\omega)/A_0$ vs. the energy $\omega$ at $\Gamma_3=0$ and the
dimensionless differential conductance $G_3/G_0$ of the third lead
vs. the voltage $V_3$ at $\Gamma_3=0.01$ for different
$\overline{\varepsilon}_0$, with $u_L=-u_R=3$. The other parameters
are same with Fig.2. Notice that the three dotted curves almost
overlap perfectly with the three solid curves so that they almost
cannot be seen in the figure.}
\end{figure}

\begin{figure}
\includegraphics[bb=-1mm 9mm 121mm 108mm, width=8cm,totalheight=6.5cm,clip=]{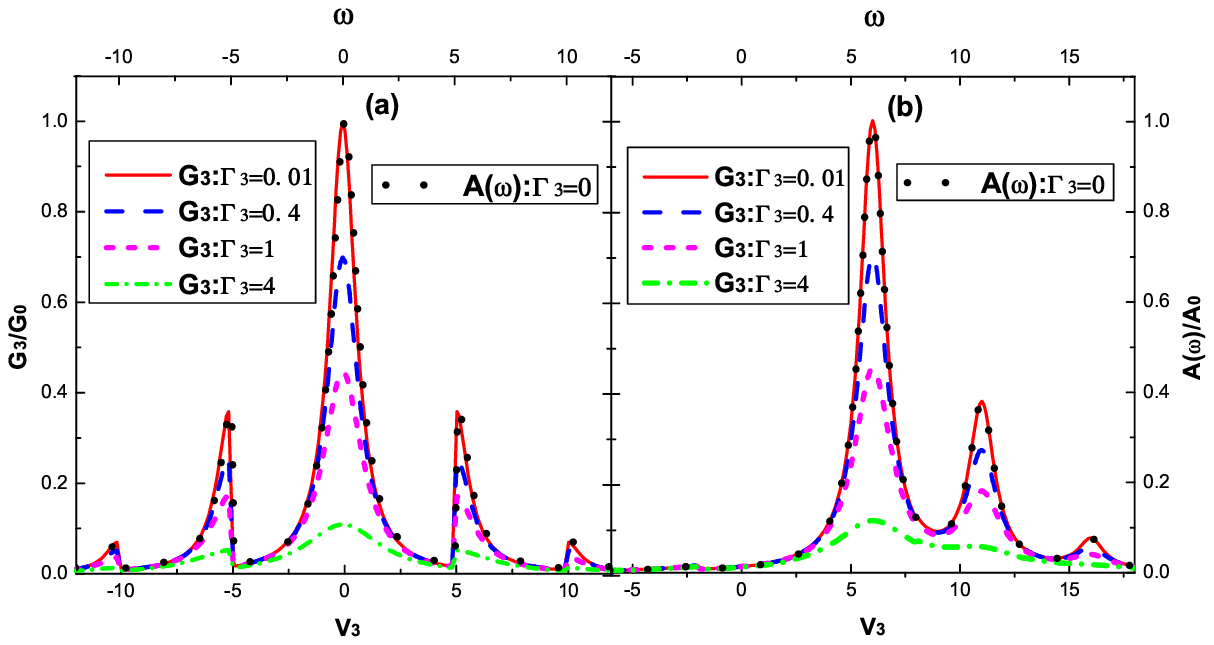}
\includegraphics[bb=-1mm 9mm 121mm 108mm, width=8cm,totalheight=6.5cm,clip=]{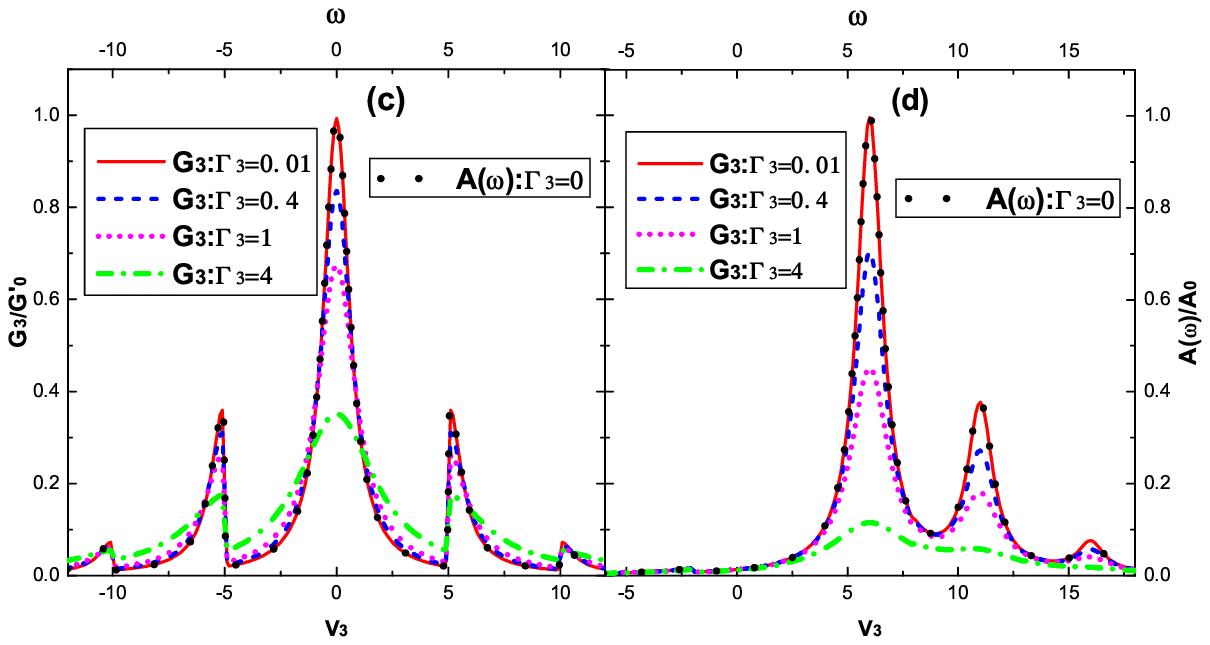}
\caption{(Color online) The dimensionless spectral function
$A(\omega)/A_0$ vs. the energy $\omega$ and the dimensionless
differential conductance $G_3$ of the third lead vs. the voltage
$V_3$ for different coupling strengths $\Gamma_3$. The parameters
are taken as: $\omega_0=5$, $T=0.02$, and $\lambda=3$. (a,c) and
(b,d) are corresponding to $u_L=u_R=0$ and
$\overline{\varepsilon}_0=0$, and $u_L=-u_R=3$ and
$\overline{\varepsilon}_0=6$, respectively. The unites $A_0
=2/\Gamma$, $G_0 =(2e^2/\hbar)(\Gamma_3/\Gamma)$, and $G'_0$ is
determined by the equation
$\int\frac{dV_3}{2\pi}\frac{G_3(V_3)}{G'_0}=\int\frac{d\omega}{2\pi}\frac{A(\omega)}{A_0}$.}
\end{figure}

\begin{figure}
\includegraphics[bb=-1mm 9mm 141mm 128mm, width=8cm,totalheight=6.5cm,clip=]{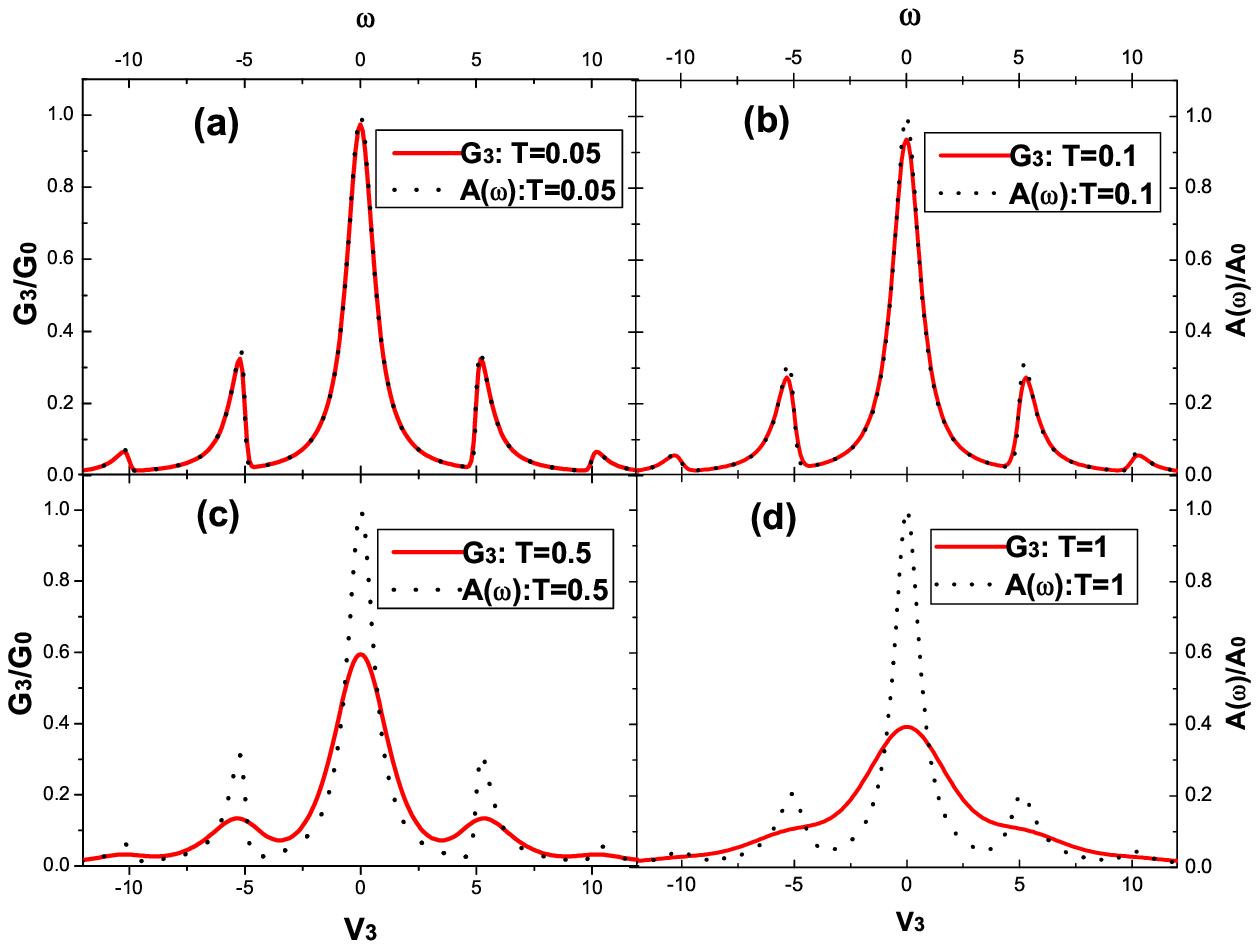}
\caption{(Color online) The dimensionless spectral function
$A(\omega)/A_0$ vs. the energy $\omega$ at $\Gamma_3=0$ and the
dimensionless differential conductance $G_3/G_0$ of the third lead
vs. the voltage $V_3$ at $\Gamma_3=0.01$ for different temperatures.
The other parameters are $u_L=u_R=0$, $\omega_0=5$,
$\overline{\varepsilon}_0=0$, and $\lambda=3$. The unites
$A_0=2/\Gamma$ and $G_0=(2e^2/\hbar)(\Gamma_3/\Gamma)$.}
\end{figure}


\begin{references}
\bibitem[*]{} Electronic address: sunqf@aphy.iphy.ac.cn

\bibitem{H. Park}
H. Park, J. Park, A. K. L. Lim, E. H. Anderson, A. P. Alivisatos,
and P. L. McEuen,  Nature (London) {\bf 407}, 57 (2000).

\bibitem{B. J. Leroy1}
B. J. LeRoy, S. G. Lemay, J. Kong, and C. Dekker, Nature {\bf 432},
371 (2004).

\bibitem{B. J. Leroy2}
B. J. LeRoy, J. Kong, V. K.Pahilwani, C. Dekker, and S. G. Lemay,
Phys. Rev. B {\bf 72}, 075413 (2005).

\bibitem{S. Sapmaz}
S. Sapmaz, P. Jarillo-Herrero, Y. M. Blanter, C. Dekker, and H. S.
J. van der Zant, Phys. Rev. Lett. {\bf 96}, 026801 (2006).

\bibitem{N. S. Wingreen}
N. S. Wingreen, K. W. Jacobsen, and J. W. Wilkins, Phys. Rev. B
{\bf 40}, 11834 (1989).

\bibitem{M. Keil}
M. Keil and H. Schoeller. Phys. Rev. B {\bf 66}, 155314 (2002).

\bibitem{D. M.-T. Kuo}
David M.-T. Kuo and Y. C. Chang, Phys. Rev. B {\bf 66}, 085311
(2002).

\bibitem{J. X. Zhu}
J. X. Zhu and A. V. Balatsky, Phys. Rev. B {\bf 67}, 165326
(2003).

\bibitem{K. Flensberg}
K. Flensberg, Phys. Rev. B {\bf 68}, 205323 (2003).

\bibitem{Z. Z. Chen}
Z. Z. Chen, R. Lu and B. F. Zhu, Phys. Rev. B {\bf 71}, 165324
(2005).

\bibitem{A. Mitra}
A. Mitra, I. Aleiner and A. J. Millis, Phys. Rev. B {\bf 69},
245302 (2004).

\bibitem{Kondo1}
Q.-F. Sun and H. Guo, Phys. Rev. B {\bf 64}, 153306 (2001); E.
Lebanon and A. Schiller, Phys. Rev. B {\bf 65}, 035308 (2002).

\bibitem{Kondo2}
S. De Franceschi, R. Hanson, W. G. van der Wiel, J. M. Elzerman,
J. J. Wijpkema, T. Fujisawa, S. Tarucha, and L. P. Kouwenhoven,
Phys. Rev. Lett. {\bf 89}, 156801 (2002); R. Leturcq, L. Schmid,
K. Ensslin, Y. Meir, D. C. Driscoll, and A. C. Gossard, Phys. Rev.
Lett. {\bf 95}, 126603 (2005).

\bibitem{G. D. Mahan} G. D. Mahan,
\begin{itshape}
 Many-Particle Physics
\end{itshape}, 3rd ed. (Plenum Press. New York, 2000).

\bibitem{Meir}
Y. Meir and N. S. Wingreen, Phys. Rev. Lett. {\bf 68}, 2512
(1992).

\bibitem{A. P. Jauho}
A. -P. Jauho, N. S. Wingreen and Y. Meir, Phys. Rev. B {\bf 50},
5528 (1994).

\bibitem{H. Huag} H. Huag and A.-P. Jauho,
\begin{itshape}
Quantum Kinetics in Transport and Optics of Semiconductors
\end{itshape}, edited by Dr. -Ing. Helmut K. V. Lotsch (Springer-Verlag, Berlin Heidelberg, 1996).

\bibitem{addnote}
Here we assume that the isolated phonon system is in equilibrium,
thus, this temperature is the same as that for the electronic part.
In fact, even if the phonon and electron temperatures are different,
namely the electron and the phonon systems are not in equilibrium,
which may occurs in a real experimental under a bias, the main
results, {\it i.e.}, the curve $A(\omega)$-$\omega$ can map into the
curve $G_3$-$V_3$, still remains, as soon as the electron
temperature is sufficiently low.

\end{references}
\end{document}